# Dirac Surface States and Nonlocal Quantum Tunneling in Topological Semiconductor $Mo_2SeTe_3$ for High-Performance Tunnel FETs

Zafar Sadik Mehrub[1], Suvodip Kundu Arnob[2], Md. Tareq Mahmud[1], Nazmul Hasan[3,†], and Alamgir Kabir[1,†]

[1]Department of Physics, University of Dhaka, Dhaka 1000, Bangladesh
[2]Department of Electrical and Electronic Engineering, Khulna University of Engineering & Technology (KUET), Khulna 9203, Bangladesh
[3]Department of Electrical and Computer Engineering, University of Rochester, NY 14620, USA

[†]Corresponding authors: nhasan5@ur.rochester.edu, alamgir.kabir@du.ac.bd

**Abstract**

A first-principles and device-level study of the quasi-two-dimensional transition-metal chalcogenide $Mo_2SeTe_3$ is performed. The material is found to be a weak topological semiconductor with a finite bulk band gap and symmetry-protected Dirac surface state, has a strong potential for next-generation low-power quantum electronic devices. An SOC-driven band inversion accompanied by an indirect semiconducting gap of ~0.75 eV is observed. Topological nontriviality is rigorously confirmed through Wannier charge-center evolution and $\mathbb{Z}2$ invariant analysis, yielding weak topological indices of (0;001), while iterative Green's-function surface-state computations corroborate Dirac-cone conducting states traversing the bulk gap on symmetry-preserving surfaces. $Mo_2SeTe_3$ additionally exhibits exceptional dynamical and mechanical stability, pronounced optical anisotropy, high dielectric polarizability, broad infrared-to-visible optical absorption, a large static dielectric constant, and substantial birefringence favorable for photonic and optoelectronic applications. Thermoelectric transport analyses further reveal enhanced carrier mobility and a competitive value of figure-of-merit under n-type doping near room temperature. To translate these material properties into device functionality, $Mo_2SeTe_3$ was implemented as the channel in a dual-source tunnel field-effect transistor (TFET) via TCAD simulations with nonlocal band-to-band tunneling, yielding subthreshold switching below the thermionic limit, a high ON/OFF current ratio, and amplified tunneling efficiency driven by SOC-induced orbital hybridization and topologically enhanced interband coupling. The concurrent realization of nontrivial bulk-boundary correspondence, robust transport, and steep-slope switching characteristics firmly establishes $Mo_2SeTe_3$ as a multifunctional quantum material platform for topological electronics, spintronics, optoelectronics, and next-generation energy-efficient nanoelectronic devices.

## Introduction

Topological semiconductors are quantum materials combining a bulk band gap with symmetry-protected Dirac-like surface states, characterized by spin–momentum locking, suppressed backscattering, and electronic properties shaped by nontrivial topology rather than conventional electronic band dispersion. These unique properties render topological semiconductors highly promising for applications in spintronics, low-power dissipation electronics, and quantum transport devices, where precise control over both charge and spin degrees of freedom is critical [1], [2], [3], [4], [5]. Fundamentally, the $\mathbb{Z}2$ topological invariant underpins the classification of time-reversal invariant systems, distinguishing trivial insulators from topologically nontrivial ones [6]. In three dimensions, strong topological insulators ($\nu0 = 1$) exhibit robust Dirac surface states on all crystallographic facets, while weak topological insulators ($\nu0 = 0$ with nonzero weak indices) host such states only on surfaces defined by lattice translational symmetry [6], [7], [8]. Emerging as a focus of recent studies, weak topological semiconductors have garnered increasing attention for their anisotropic surface transport properties, tunable topological responses, and potential for integration with crystallographic engineering in advanced device architectures [9], [10], [11], [12].

A pivotal principle governing topological semiconducting behavior is spin–orbit coupling (SOC), which, in materials containing heavy elements splits the bands near the Fermi level and especially for topologically nontrivial properties SOC induces a band inversion between the valence and conduction bands or two adjacent bands in the vicinity of the Fermi level, an effect that can be elucidated through the modulation of the chemical potential. This band inversion leads to the formation of Dirac surface states that span the bulk gap and remain protected under time-reversal symmetry, thereby establishing a nontrivial band topology [13], [14], [15]. The presence of a finite semiconducting band gap, as opposed to a metallic bulk, is essential for facilitating electrostatic gating, carrier modulation, and efficient integration with conventional room-temperature semiconductor technologies [16], [17], [18], [19]. Many topological semiconductors exhibit diverse physical properties beyond their electronic topology that are highly advantageous for next-generation electronic and energy device applications. These attributes enable topological semiconductors to serve not only as efficient spin-transport channels but also as active materials in optoelectronic, photonic, improved charge storage, and energy-harvesting devices [20]. Moreover, the mechanical robustness of bulk topological semiconductors makes them particularly well-suited for thin-film deposition, heterostructure fabrication, and scalable device integration, paving the way for advanced technological platforms like superconducting spintronics [21], [22], [23], [24], [25].

Significant advancements have been made in the field of topological quantum materials; however, a critical challenge persists in achieving intrinsic, tunable systems where symmetry breaking and band topology can be precisely controlled without relying on the alignment of twisted heterostructure interfaces. While materials such as $MoTe_2$ exhibit Weyl semimetal states and demonstrate how lattice modifications can induce topological phase transitions, their sensitivity to phase variations and defect-induced inconsistencies limit reproducibility and hence their integration into devices. In this context, layered transition-metal chalcogenides have emerged as a promising class of materials for developing topological semiconductors. These materials, characterized by electronic states derived from d-orbitals and paired with heavy chalcogen elements like Te, benefit from enhanced spin–orbit coupling (SOC) while maintaining semiconducting band gaps [26], [27], [28], [29].

Besides, continuous scaling of conventional metal–oxide–semiconductor field-effect transistors (MOSFETs) has necessitated the development of advanced device engineering to overcome the 60-mV/dec subthreshold slope (SS) limit at room temperature. To enable energy-efficient electronics, devices capable of operating below the thermionic limits of traditional MOSFETs are critical, where Tunnel FETs (TFETs) are particularly promising due to their reliance on band-to-band tunneling for current injection over thermal emission [30], [31], [32]. It is now well developed in demonstrating that Weyl physics can be realized in the elemental semiconductor tellurium (Te) [33]. These "Weyl semiconductors" exhibit Weyl band crossings near the valence band maximum and conduction band minimum, integrating nontrivial band topology with the presence of a bandgap. This unique combination facilitates the simultaneous manipulation of charge transport and chirality-related topological transport, presenting opportunities for the development of advanced topological devices, such as topological field-effect transistors (TFETs), capable of concurrently toggling conducting and topological states [33], [34]. However, achieving steep switching and high ON-state current in scaled TFETs remain challenging, as defects, interface traps, and nonideal tunneling junctions limit performance benchmarks.

Topological semiconductors offer a compelling advancement in that context, with their edge or surface states facilitating high-speed, low-loss transport while exhibiting strong resilience to backscattering and disorder. This phenomenon has been observed in two-dimensional topological-insulator ribbons and topological stanene nanoribbons, where edge conduction exhibits remarkable robustness against vacancies and structural defects [35], [36], [37]. Recent investigations have shown that electrostatic gating can transition a topological semiconductor from Weyl-like transport to a state within the bandgap, enabling substantial conductivity modulation and low-voltage topological phase-change switching with a high ON/OFF ratio [38]. Moreover, topological-insulator tunneling structures have exhibited pronounced quantum-switching behavior [39]. This makes the integration of topological semiconductors as channel materials in TFETs a promising strategy for developing ultralow-power, highly scaled, and faster-switching nanoelectronic devices [19], [40], [41], [42].

Addressing the critical intersection of materials and device engineering, this study presents a comprehensive first-principles investigation of a novel quasi-two-dimensional transition metal chalcogenide compound, $Mo_2SeTe_3$, composed of MoTe2 and MoTeSe. The analysis identifies $Mo_2SeTe_3$ as a topological semiconductor with a weak topological insulating phase, where spin–orbit coupling induces band inversion and a finite bulk bandgap, confirmed as topologically nontrivial (($\mathbb{Z}2$) indices: 0; 0 0 1) with Dirac-like surface states crossing the gap on symmetry-preserving facets. We also investigate the optical, elastic, and thermoelectric properties of $Mo_2SeTe_3$ to assess its suitability for multifunctional device applications. The results identify $Mo_2SeTe_3$ as a previously unexplored topological semiconductor that uniquely integrates a finite semiconducting bandgap with Dirac surface states, alongside favorable optical and transport characteristics. Furthermore, the material is characterized as a channel material within a TFET, serving as a proof-of-concept for its potential application in advanced electronic devices. This combination establishes $Mo_2SeTe_3$ as a promising material platform for applications in spintronics, optoelectronics, and advanced quantum electronic devices.

## Materials System and Computational Methodology

$Mo_2SeTe_3$ crystallizes in a trigonal lattice with space group *P3m1* (No. 156), forming a layered configuration with atomic positions defined by Wyckoff coordinates. The optimized lattice constants are a = b = 3.494 Å and c = 14.763 Å, with interaxial angles $\alpha = \beta = 90°$ and $\gamma = 120°$, consistent with trigonal symmetry. The crystal structure is presented in Fig. 1(a).

The structural, electronic, mechanical, and optical properties of $Mo_2SeTe_3$ were analyzed using spin-polarized density functional theory (DFT) within the framework of the Vienna Ab-initio Simulation Package (VASP) [43], [44], [45]. The Generalized Gradient Approximation (GGA) with the Perdew–Burke–Ernzerhof (PBE) functional was employed for exchange-correlation effects [46], [47], and Projector Augmented Wave (PAW) pseudopotentials were used to model electron–core interactions [48] .

Phonon properties were evaluated using Density Functional Perturbation Theory (DFPT), and topological aspects were studied through Wannier-based analysis, confirming non-trivial surface states. Thermoelectric and transport parameters were determined by solving the Boltzmann Transport Equation, considering temperature-dependent variations.

A detailed account of computational parameters and convergence criteria is provided in the Supplementary Information (Section 1).

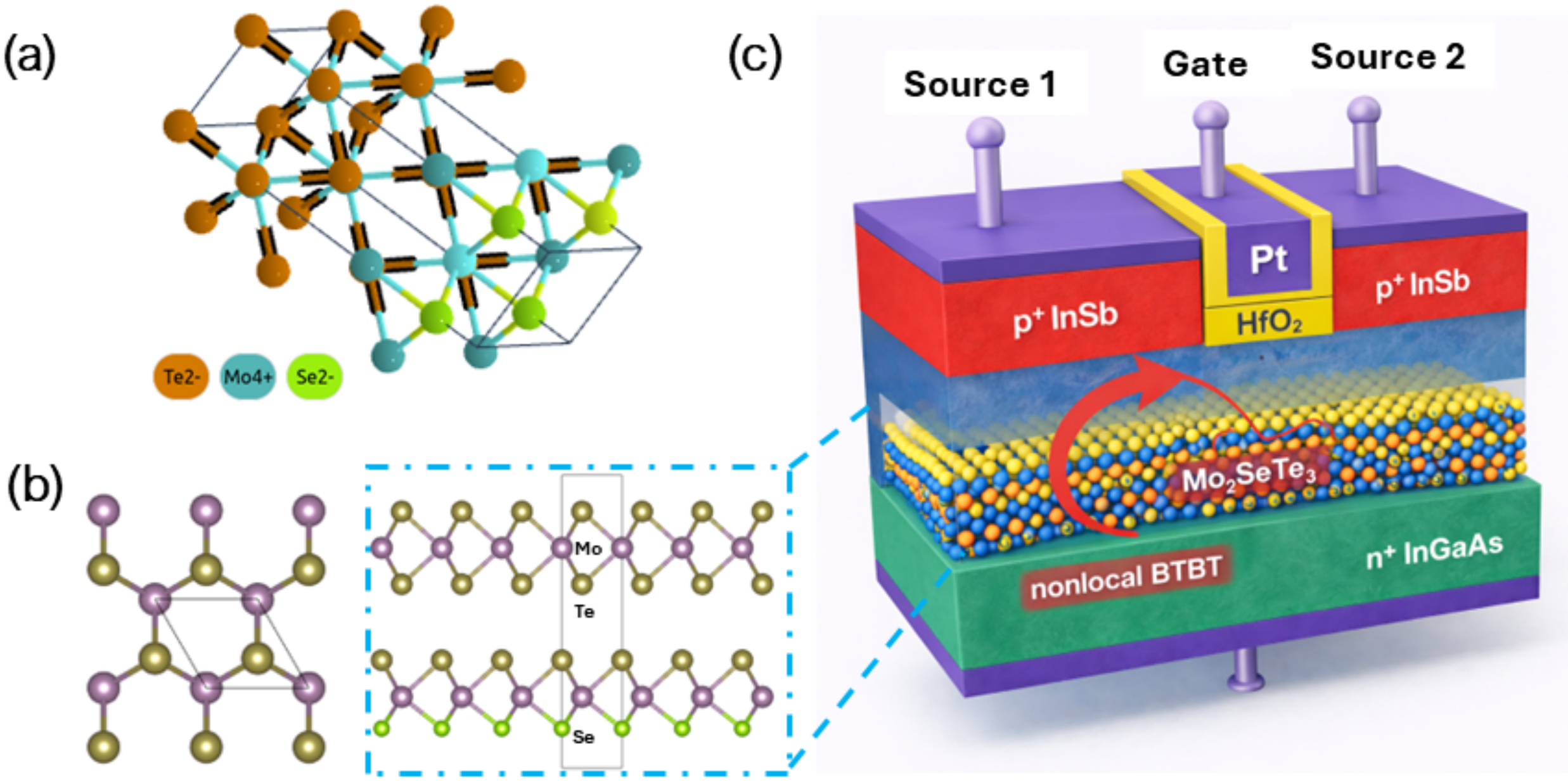

Figure 1: Crystal of $Mo_2SeTe_3$ material system in (a), unit cell representation with geometry optimized structure in (b), and modelled tunnel field-effect transistor (TFET) using $Mo_2SeTe_3$ as channel material in (c).

## TCAD Device Electrostatics for Tunneling FET

### Device Architecture:

Fig. 1(c) shows the proposed dual-source TFET comprising two symmetrically positioned $p^+$ InSb source regions, a lightly n-doped $Mo_2SeTe_3$ channel, and an $n^+$ InGaAs drain. The gate stack consists of a Pt metal gate with a work function of 4.9 eV and a 2 nm amorphous $HfO_2$ gate dielectric. The device is designed with a gate length of 10 nm and a channel length of 50 nm, where the gate and channel lengths correspond to the gate contact length and the separation between source and drain contacts, respectively. The source, channel, and drain regions are doped to $1\times10^{20}$ $cm^{-3}$, $1\times10^{15}$ $cm^{-3}$, and $1\times10^{19}$ $cm^{-3}$, respectively. In contrast to a conventional single-source TFET, the present design incorporates two source regions flanking the recessed gate, thereby creating dual tunneling junctions and increasing the effective tunneling perimeter [49], [50]. This dual-source configuration enhances electric-field concentration near the source/channel interfaces, improves gate-to-junction electrostatic coupling, and increases the available carrier injection area, leading to improved ON-state current while maintaining low OFF-state leakage. The recessed gate geometry further strengthens band modulation in the channel region, making the proposed structure attractive for high-performance low-power tunneling devices [51].

**Numerical Modeling:**

The proposed dual-source InSb/ $Mo_2SeTe_3$/ InGaAs heterostructure vertical tunnel field-effect transistor (DS-VTFET) was investigated using the ATLAS module of the Silvaco TCAD simulator in a two-dimensional framework. The device operation is governed by the coupled interaction of electrostatics, quantum confinement and nonlocal band-to-band tunneling. To capture these effects self-consistently, the simulation framework solves the semiconductor transport equations together with the Schrödinger equation, Poisson equation and nonlocal band-to-band tunneling (BTBT) model [52], [53]. Detailed of the numerical modeling of all formalisms are noted in supplementary information, section 2.

## Quantum Electronic Structure and Device-level Implications (Materials Physics and Device Characteristics)

### Phonon Dispersion, Dynamic and Mechanical Stability

The phonon dispersion curve and Phonon-DOS of $Mo_2SeTe_3$, shown in Fig. 2(a), confirm its dynamical stability, as there are no negative or imaginary phonon frequencies throughout the entire Brillouin zone. For $Mo_2SeTe_3$, which has 6 atoms per unit cell, a total of 18 phonon bands are observed, as each atom contributes 3 vibrational degrees of freedom. Here 3 acoustic bands approach zero near the high-symmetry Γ point and 15 bands with non-zero frequencies are observed as optical branch. The bands are sometimes degenerate at high symmetry points so they may appear as overlapping bands in the phonon dispersion curve within the Brillouin zone. However, the optical modes are clustered into distinct frequency ranges (low-optical 3.3–4.7 THz, a mid-set 5.6–6.5 THz, and higher modes up to 9 THz), with partial separation between acoustic and optical modes influenced by sublattice mass differences. This separation can limit certain scattering processes, while overlapping low-frequency optical and acoustic modes enhance hybridization and scattering under suitable conditions [54]. In the right panel of Fig. 2(a), phonon density of states (DOS) displays prominent peaks near 3.5 - 4 THz and 6 THz, along with a broader distribution in the 7-9 THz range. Sharp DOS features arise from flat regions in the phonon dispersion with low group velocities, where many q-points contribute at similar frequencies (ω), known as the phononic van Hove singularity, based on the harmonic approximation for phonons [55]. Low-energy optical phonons (3-5 THz) enable inelastic channels for phonon-assisted tunneling in tunneling-FETs, affecting current onset and temperature dependence [56]. In topological materials, ℤ2 indices signify protected boundary states, while Fermi-arc contours define Weyl/Dirac semimetals, both rooted in bulk-boundary correspondence [6]. Thus, a stable phonon spectrum is crucial for interpreting topological properties and modeling devices that incorporate phonon effects.

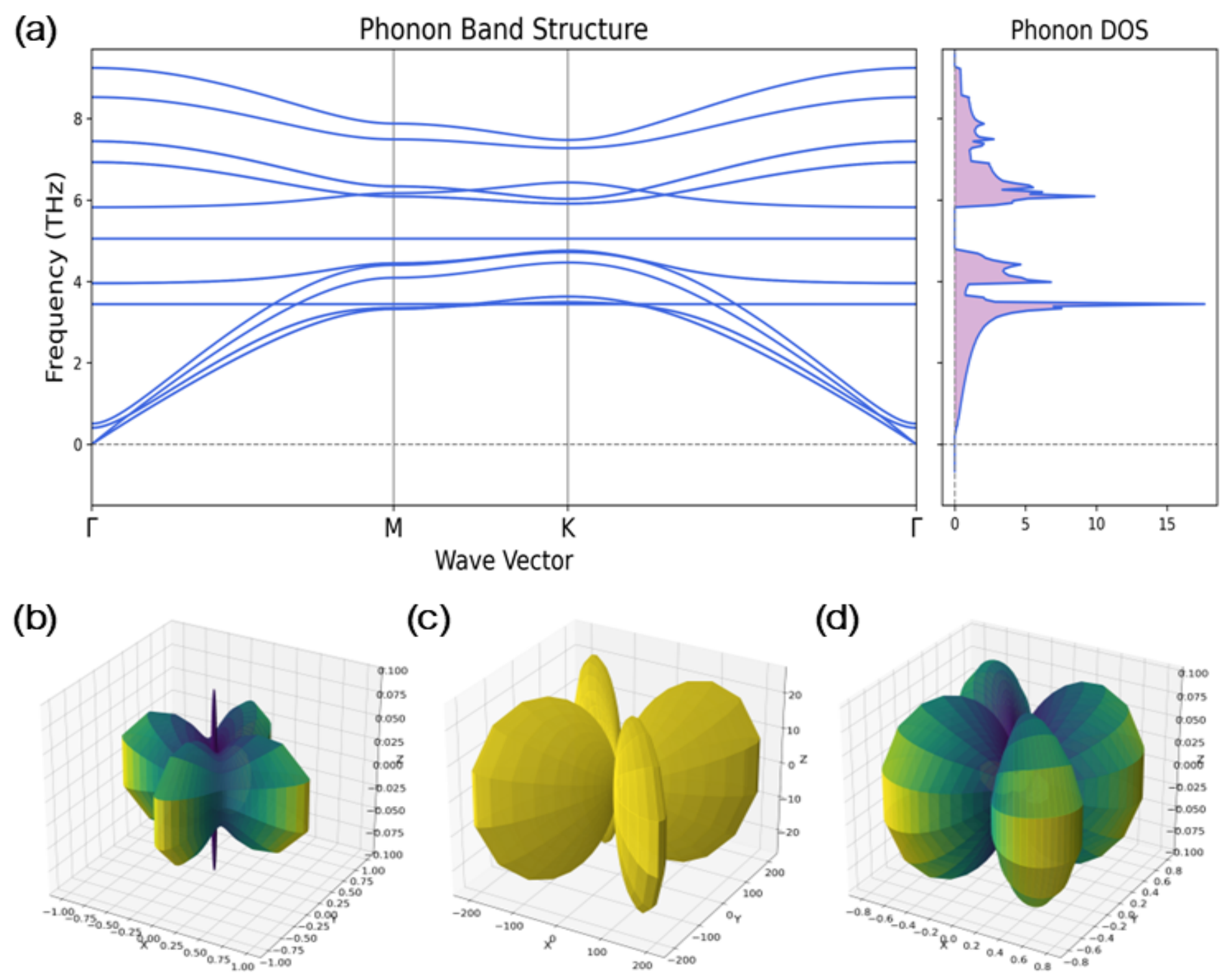

Figure 2: Stability Analysis of $Mo_2SeTe_3$ Materials – (a) Phonon dispersions and phonon density of states (Phonon-DOS) illustrating the dynamic stability of the material; (b)-(d) Mechanical stability parameters including Young's modulus, Poisson's ratio, and bulk modulus.

The detailed investigation of the mechanical and elastic properties of $Mo_2SeTe_3$ provides valuable insights into its structural stability, bonding characteristics, and potential applications. These properties are essential for optimizing material selection, enhancing fabrication techniques, and predicting material behavior under operational conditions. Elastic constants and moduli were determined based on Born's stability criteria [57]; $Mo_2$SeTe3 satisfies all the required conditions, confirming its mechanical stability. To further explore the material's dynamic properties, parameters such as Cauchy pressure, Pugh's ratio, and Debye temperature are presented in table S3 in the supplementary informations. The bulk modulus measures a material's resistance to volumetric deformation under pressure, and $Mo_2SeTe_3$ exhibits a relatively high bulk modulus of 36.22, indicating its flexibility and ability to undergo significant volume changes under pressure. Additionally, the material's high Young's modulus highlights its robustness under pressure, making it suitable for applications such as thin-film fabrication and energy devices [58], [59]. The material also demonstrates a high shear modulus (G), which quantifies resistance to shear deformation, making it ideal for applications in flexible electronics. Furthermore, the Poisson's ratio ($\sigma$) of $Mo_2SeTe_3$ reveals a high anisotropy ratio of 19.008, indicating significant variation in the material's response to stress, as illustrated in Fig. 2(c). Similar trends are observed for the anisotropic Young's modulus and shear modulus. The Kleinman parameter, which evaluates internal strain within materials and ranges from 0 to 1 for solids, is calculated to be 0.43 for $Mo_2SeTe_3$, further emphasizing its structural stability.

**Electronic Band Structure**

The electronic band structure of $Mo_2SeTe_3$ along the high-symmetry points in the Brillouin zone of the hexagonal lattice is depicted in Fig. 3(a) and 3(c), illustrating the material's behavior without and with the influence of spin-orbit coupling (SOC), respectively. The band structure reveals the semi-conducting nature intrinsically, as no bands cross the Fermi level, and the material has a band gap of Eg = 0.7507 eV between the Valence Band Maximum (VBM) and the Conduction Band Minimum (CBM). The impact of SOC on the band structure is particularly significant due to the presence of heavy metals like molybdenum (Mo), which exhibit strong spin-orbit interactions as observed from the atomic density of states in Fig. 3(b) and (d). SOC induces band splitting near high-symmetry points in the Brillouin zone, altering the electronic properties of the material. For instance, in Fig. 3(c), specific regions such as $\Gamma - M$ (from -0.5 eV to -1.5 eV), $K - \Gamma$ (from 1 eV to 2 eV), and $A - L$ (from 0.5 eV to 1.5 eV) show degenerate upper and lower bands. This splitting behavior is a direct consequence of SOC, which modifies the energy levels and introduces additional complexity to the electronic structure. Such effects are crucial for understanding the material's electronic behavior and its potential in spintronic applications, where SOC plays a pivotal role in controlling spin-dependent phenomena.

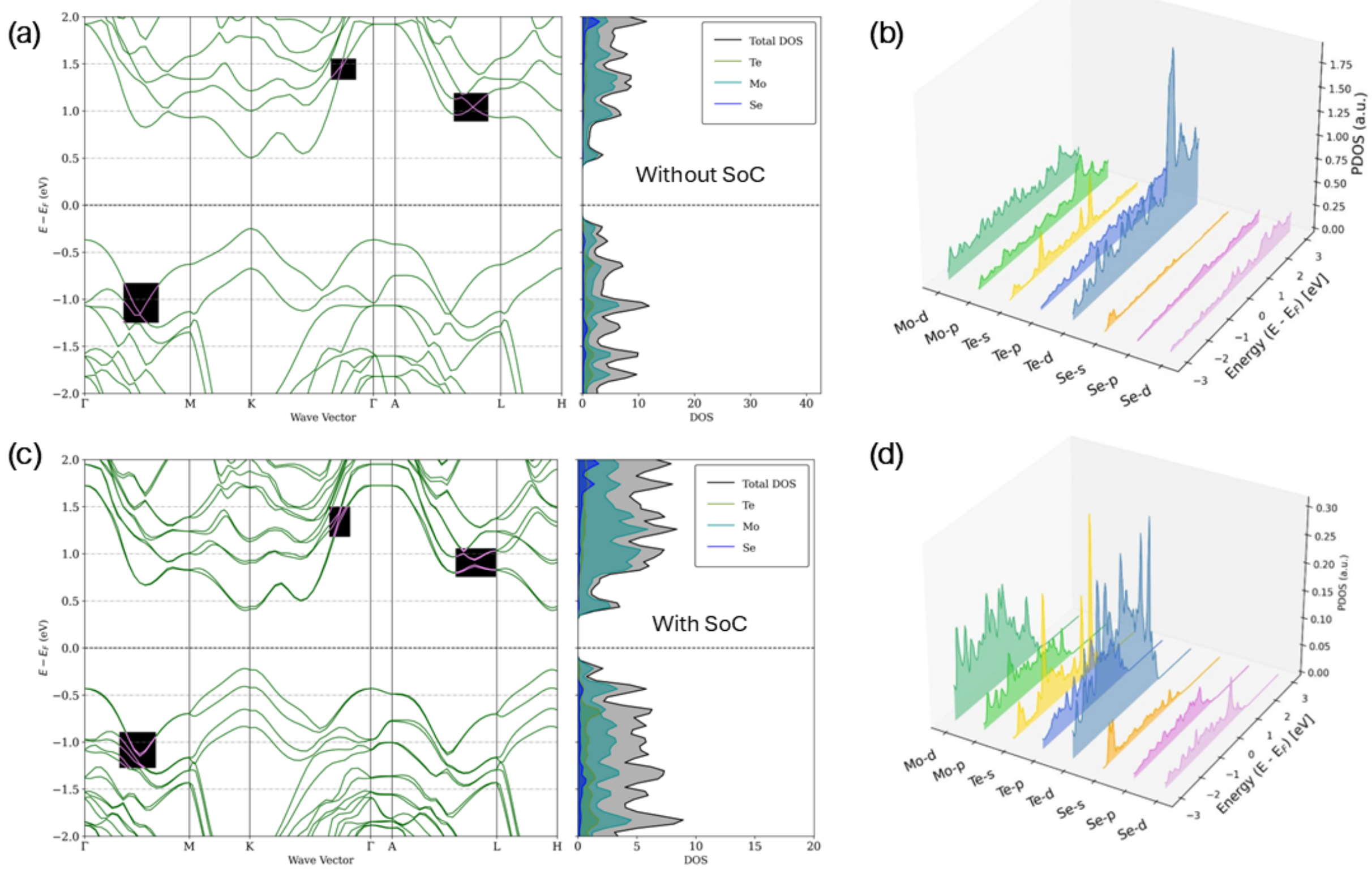

**Figure 3.** (a) Electronic band structure and atom-projected density of states (PDOS) of $Mo_2SeTe_3$ without spin–orbit coupling (SOC). (c) Corresponding band structure and PDOS with SOC. The energy window is set from −2 eV to 2 eV with respect to the Fermi level. Black markers highlight the band opening induced by SOC near the Fermi level. (b) Orbital-resolved density of states without SOC, and (d) with SOC.

When spin-orbit coupling (SOC) is accounted for, an energy gap is observed in the marked band crossings of Fig. 3(a), indicating that SOC effectively lifts the degeneracy at specific points in the band structure. The appearance of band crossings' opening and thus lifting degeneracy, as shown in Fig. 3(c), is one of the most intriguing and significant features, resembling key characteristics typically associated with topological insulators (TIs) [60]. In the intermediate energy range near the Fermi level, the contributions from molybdenum (Mo) and tellurium (Te) atoms are particularly prominent, following Fig. 3(d). To analyze the density of states (DOS) in greater detail, the orbital contributions of each atom to the total DOS near the Fermi level were evaluated. Without considering SOC, the *d* orbitals of Mo and Te dominate the near Fermi energy, with additional contributions from the *p* orbitals of Te and selenium (Se). However, when SOC is applied, the *d* orbital of Te becomes the most dominant, surpassing other atomic orbitals. This shift highlights the significant influence of SOC on the electronic structure and orbital interactions within the material.

## Topological Phase of Electronic Structure

Topological insulators exhibit a characteristic band inversion in their bulk phase, where a significant spin-orbit coupling (SOC) induces the opening of an energy gap. This is accompanied by topologically protected surface states at the non-trivial or topological interface junctions.

Similar to quantum spin hall layers, these states are safeguarded by time-reversal symmetry (TRS) and are characterized by a $\mathbb{Z}2$ topological invariant [12]. Additionally, self-consistent field calculations for $Mo_2SeTe_3$ reveal that the total magnetization converges to zero, confirming that the material is non-magnetic. The combination of zero magnetization and preserved TRS provides a strong basis for calculating the $\mathbb{Z}2$ topological invariants to classify the system's topological nature.

To determine the topological classification, the $\mathbb{Z}2$ topological indices (v0, v1v2v3) were calculated for six time-reversal invariant momentum planes: (a) k1 = 0.0, (b) k1 = 0.5, (c) k2 = 0.0, (d) k2 = 0.5, (e) k3 = 0.0, and (f) k3 = 0.5. Here, v0 indicates a strong topological phase if its value is 1, while the remaining three indices represent weak topological phases. These indices were determined using Wannier charge centers to visualize the Wilson loop and calculate the $\mathbb{Z}2$ invariants. The orbital contributions, with a detailed consideration of SOC, were incorporated to construct 62 maximally localized Wannier functions using a tight-binding Hamiltonian, which was validated by comparing the band structure and density of states (DOS) obtained through the Wannier90 and WannierTools package [61], [62].

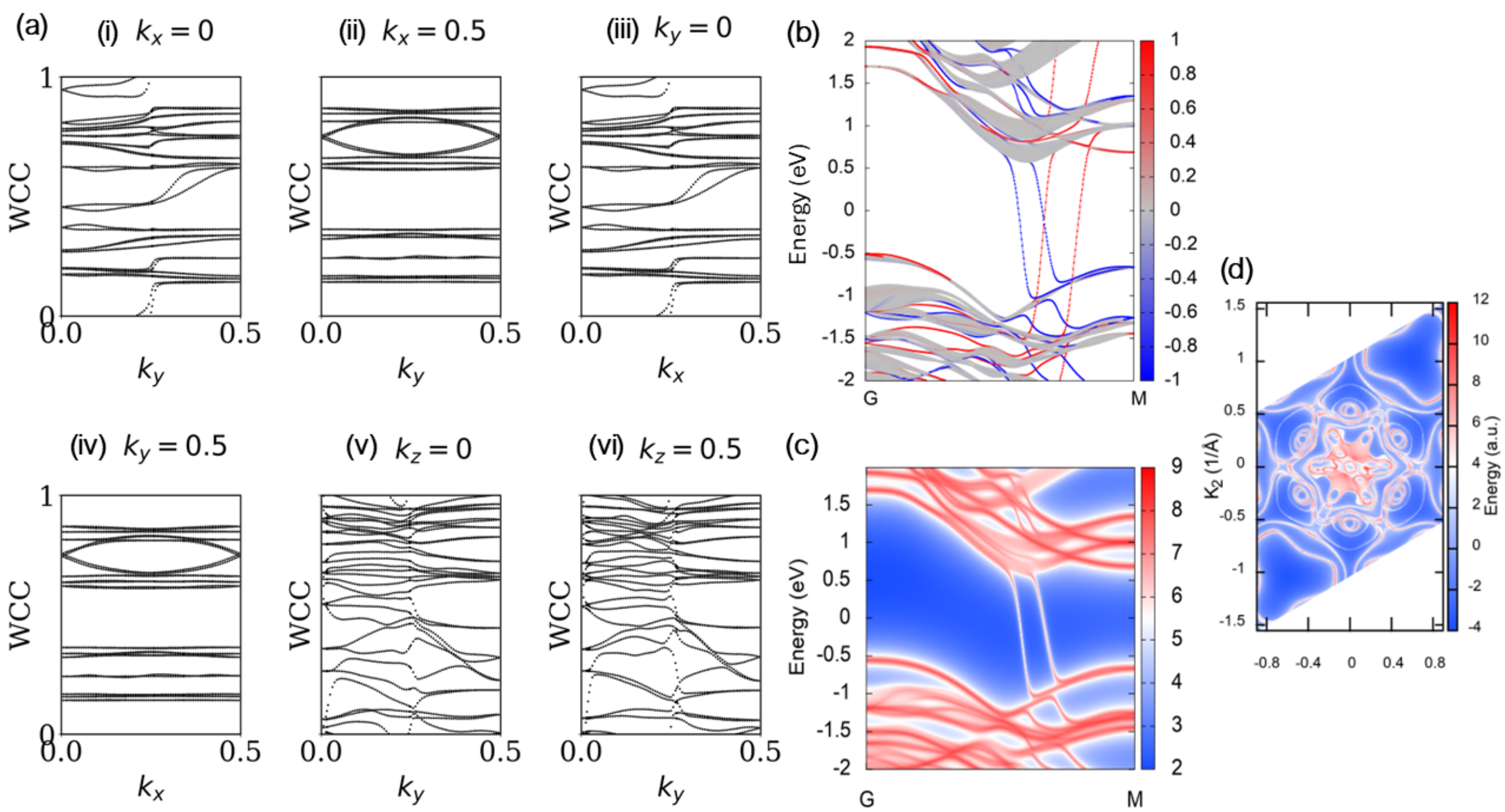

Figure 4: (a) Wannier charge centers (WCCs) loops across the six time-reversal invariant planes. Surface states obtained using the iterative Green's function method based on the Wannierized Hamiltonian along with the intensity scale. (b) Slab band showing conducting surface state, (c) Simulated ARPES spectrum along the G(Γ)–M path near the Fermi level, (d) Presence of a Fermi arc.

From the Wannier Charge Center (WCC) loop shown in Fig. 4(a), it is observed that there are odd numbers of crossings in the kz plane, specifically in both kz = 0.0 and kz = 0.5 planes. Based on this observation, the calculated topological indices are v0 = 0, v1 = 0, v2 = 0, and v3 = 1, which can be compactly expressed as (0; 001). This classification clearly identifies $Mo_2SeTe_3$ as a weak topological insulator. If the bulk system exhibits nontrivial topological properties, nontrivial conducting states will emerge on the surface [63]. Sometimes, the nontrivial conducting states are beneficial for modern electronic device modeling because it acts as an insulator in the bulk but contain gapless conducting states on the surface. As these states are "topologically protected" by time-reversal symmetry, electrons cannot be easily backscattered by impurities or defects. To scatter backward, the electron would have to flip its spin, which is forbidden unless the defects or perturbations are magnetic. So this crucial properties prohibit electron's backscattering within the device and enhance the control over the whole carrier channel resulting in an increase of the device potential [64], [65].

*Table 1. $\mathbb{Z}2$ topological invariants for six time-reversal invariant momentum planes.*

| Plane | $\mathbb{Z}2$ Index |
|---|---|
| kx = 0.0, ky-kz plane | 0 |
| kx = 0.5, ky-kz plane | 0 |
| ky = 0.0, kx-kz plane | 0 |
| ky = 0.5, kx-kz plane | 0 |
| kz = 0.0, kx-ky plane | 1 |
| kz = 0.5, kx-ky plane | 1 |

To visualize these surface states, DFT simulated Angle-Resolved Photoemission Spectroscopy (ARPES) is shown in Fig. 4(c), the valence and conduction bands exhibit conducting states at the surface, which is a hallmark of the topological phase. These conducting surface states are protected by time-reversal symmetry and remain invariant under perturbations. Furthermore, the observation of Fermi arcs reinforces the topological nature of $Mo_2SeTe_3$ and highlights its potential for Fermi arc-driven exotic properties [66].

**Optical Properties**

Materials with topological properties exhibit unidirectional spin-polarized photon transmission through robust helical surface edge states, which are resistant to disorder. Experimental studies have demonstrated this phenomenon without requiring external magnetic fields or breaking time-reversal symmetry [67], [68], [69]. Hence, topological materials, particularly gapped insulators, are emerging for their potential in photonic technologies, including lasers and optoelectronic devices [70], [71]. To elucidate the optical response and potential device applications of the material, we analyzed the dielectric properties. It was derived using the Kramers-Kronig relation for the complex dielectric function, $\varepsilon(\omega) = \varepsilon 1(\omega) + i\varepsilon 2(\omega)$, where $\varepsilon 1(\omega)$ represents the real part, which determines light polarization and dispersion, and $\varepsilon 2(\omega)$ represents the imaginary part, which describes the material's absorptive behavior [72], [73]. These components form the foundation for understanding optical parameters such as the refractive index $n(\omega)$, extinction coefficient $k(\omega)$, absorption coefficient $\alpha(\omega)$, reflectivity $R(\omega)$, and energy loss spectra $L(\omega)$. The relationships between the dielectric constants and these optical parameters are well-established in supplementary information, and DFT computed those optical behaviors are represented in Fig. 5.

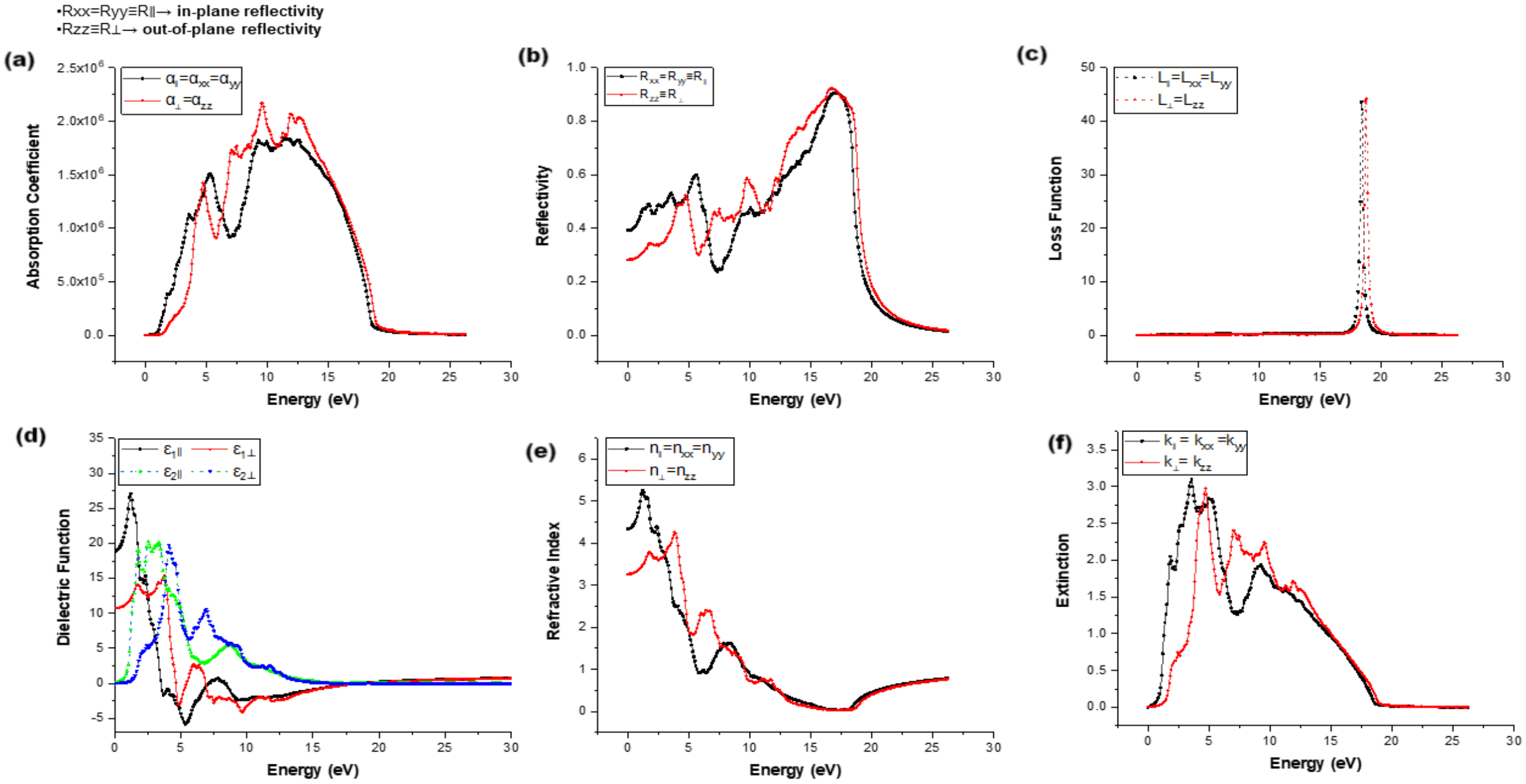

Figure 5: Optical properties of $Mo_2SeTe_3$: (a) Absorption coefficient α(ω), (b) Reflectivity R(ω), (c) Energy loss function L(ω) (d) Dielectric Function (e) Refractive Index (n) (f) Extinction. These properties show anisotropy. The figures in black show in-plane or parallel optical response, whereas the red illustrates out-of-plane or perpendicular optical properties.

The calculations span an energy range of 0 to 30 eV. A pronounced uniaxial optical anisotropy is strictly preserved across all spectra, driven by the structural asymmetry between the basal plane (xx=yy≡∥) and the perpendicular axis (zz≡⊥). For $Mo_2SeTe_3$, the static dielectric constant (ε1(0)) is 18.25, the static refractive index (n(0)) is 4.3, and the static reflectivity (R(0)) is 33.96%. A higher dielectric constant indicates increased polarizability and plays a crucial role in reducing charge carrier recombination rates, thereby enhancing the performance and efficiency of optoelectronic devices [74]. As shown in Fig. 5(d), the optical onset reveals a strong directional dependence. In the low-energy regime (1–5 eV), the in-plane component (ε2∥) exhibits a sharp, dominant absorption peak near 3 eV. Conversely, the out-of-plane component (ε2⊥) demonstrates a delayed primary excitation, peaking at roughly 7 eV. This distinct splitting in transition probabilities indicates highly anisotropic orbital hybridization near the Fermi level. Again following the same Fig. 5(d), the real part of the dielectric spectrum ε1(ω) reveals asymmetry. The in plane components increases up to an energy value of 2.235 eV, then decreases, reaching its lowest value at 5.01 eV, and remains negative up to 17 eV. In the negative region, the material cannot support propagating electromagnetic waves, causing UV rays within this energy range to be reflected, consistent with the reflection vs. energy plot from Fig. 5(b). Beyond 20 eV, ε2(ω) approaches zero, signifying transparency with negligible absorption in high-energy regions. This transparency suggests the absence of electronic transitions at these energy levels, likely due to the lack of suitable electronic states for such transitions.

In Fig. 5(e), in the visible spectrum (1.5–3 eV), the in-plane refractive index peaks dramatically at $n_\parallel \approx 5.2$, while the out-of-plane index remains lower, peaking later at $n_\perp \approx 4.2$. This massive intrinsic birefringence ($\Delta n = n_\parallel - n_\perp$) is highly desirable for the design of compact optical waveplates, phase-retarders, and polarization-sensitive photodetectors. As shown in Fig. 5(a),

The 1st peak of the absorption coefficient graph occurs around 5 eV whereas the highest local maxima is around 10 eV. Furthermore, the absorption coefficient (α) rapidly scales to the order of $10^6$. Such robust optical attenuation over a broad solar to UV spectrum suggests that ultra-thin films of this material could function as highly efficient absorber layers in photovoltaics or high-responsivity UV sensors. Beyond the interband transition regime, the material transitions to a collective excitation phase, captured by the energy loss function, which varies inversely with the imaginary part of the dielectric constant. The L(ω) spectra (Fig. 5(c)) display exceptionally sharp and distinct peaks at approximately 18.5 eV for $L\|$ and 19.0 eV for $L\perp$. These sharp peaks indicate a dominant excitation mode, commonly interpreted as bulk plasmon resonance [75]. The existence of highly localized, intense plasmon resonances in the 18–19 eV range, coupled with the material's inherent directionality makes it a compelling candidate for extreme-ultraviolet (EUV) plasmonics, directional deep UV waveguides, and high-energy electron energy-loss spectroscopy (EELS) calibration standards.

### Thermoelectric and Transport Properties

Thermoelectric properties of topological insulators have gained significant attention due to shared characteristics with thermoelectric materials, such as narrow band gaps and heavy metal composition[76], [77]. This study focuses on identifying key thermoelectric properties, particularly the Seebeck coefficient, thermal and electrical conductivity, power factor, and figure of merit as a function of temperature ranges from 50K to 300K. The thermodynamic stability of the system across the evaluated temperature range is confirmed by the continuous, monotonic evolution of the free energy, entropy, and heat capacity ($C_v$) [Fig. 6(f)], with $C_v$ approaching the classical Dulong-Petit limit at elevated temperatures.

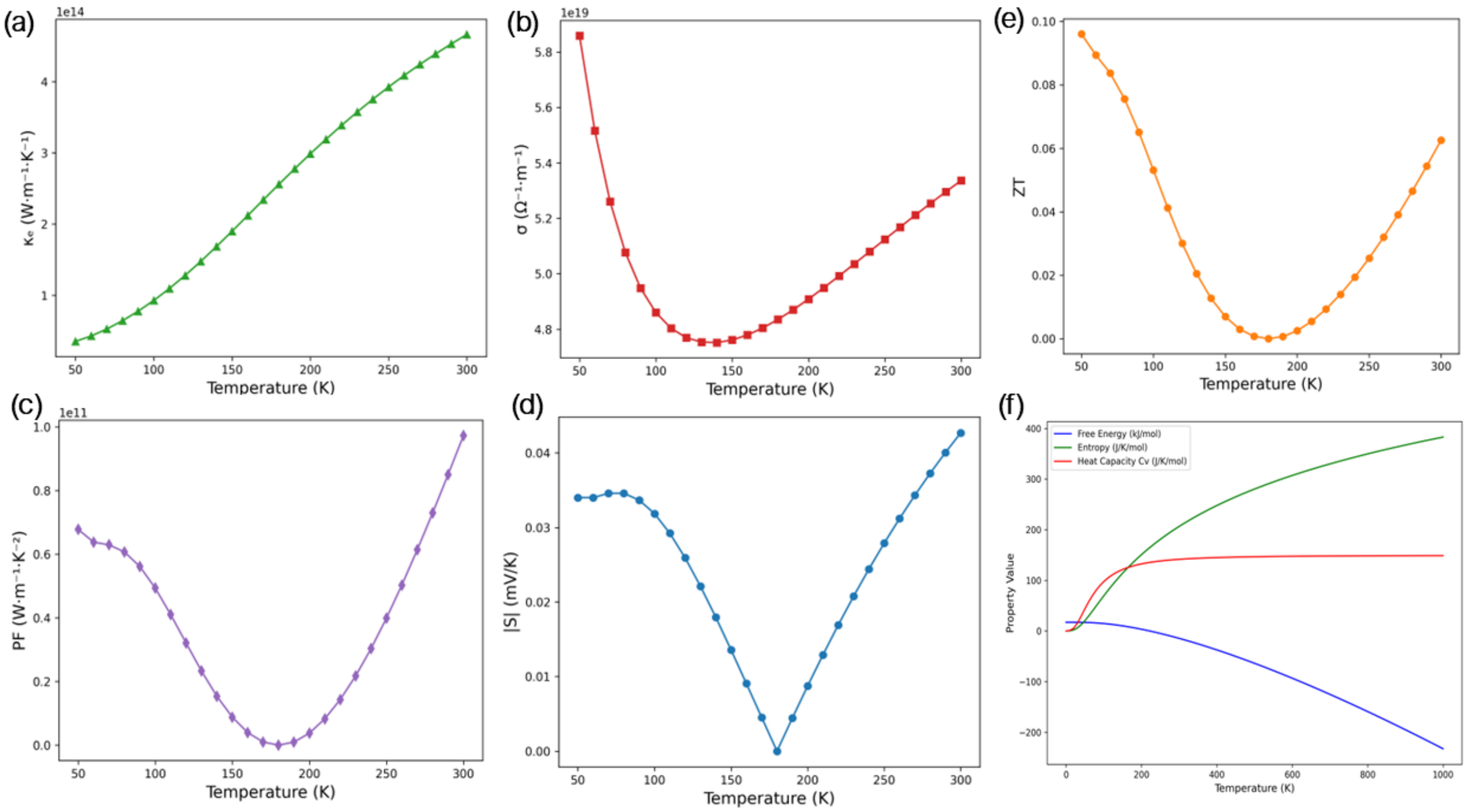

Figure 6: Temperature dependence of key thermoelectric properties: (a) Electronic thermal conductivity, (b) Electrical conductivity, (c) Power factor, and (d) Seebeck coefficient. (e) Figure of Merit ZT (f) Free energy in Blue, Heat capacity in Red, Entropy in Green.

As shown in Fig. 6(a), thermal conductivity increases monotonically with temperature, which is advantageous for thermoelectric efficiency. The calculation of thermal conductivity was evaluated based on electronic contributions, without accounting for the phononic contribution due to inherent limitations of the BoltzTraP2 code. Conversely, the electrical conductivity decreases across nearly half of the investigated temperature range [Fig. 6(b)], attaining its peak value of $5.8 \times 10^{19}\ \Omega^{-1}\ m^{-1}$ (per relaxation time) in the low-temperature domain. Following to the same figure the curve has local minima at 150 K then subsequently increases up to the ambient temperature of 300K. As seen from Fig. 6(c), the power factor ($PF=\sigma S^2$, where S is the Seebeck coefficient) exhibits similar properties and has a prominent minimum at this exact temperature. Conversely, Fig. 6(d) illustrates the temperature-dependent behavior of the Seebeck coefficient in the undoped situation, which is notably high at very low temperatures then drops sharply after 150K and vanishes at 180K. This zero-crossing is a classic signature of the bipolar effect [78]. Here the total Seebeck coefficient ($S_{tot}$) can be defined as the weighted average of the contributions from electrons ($S_e$) and holes ($S_h$),

$$S_{tot} = \frac{(\sigma e * Se \ + \ \sigma h * Sh)}{(\sigma e \ + \ \sigma h)}$$

So at a temperature of 180K, the contributions of both electrons and holes to the transport coefficient are equivalent ($S_e$ & $S_h$), resulting in a net Seebeck coefficient of zero. As it seen from the Fig. 6(d) the Seebeck coefficient is again higher near room temperature. This trend is favorable for generating thermoelectric voltage both at low and near-room temperatures. Finally, following Fig. 6(e), the intrinsic figure of merit, ZT, remains negligible (<0.1) across the entire temperature range, thereby emphasizing the imperative for extrinsic doping to enhance the material's potential as a thermoelectric generator and break the electron-hole symmetry. Additionally, the thermoelectric properties of $Mo_2SeTe_3$ were further analyzed under electron and hole doping conditions and illustrated in Fig. 7.

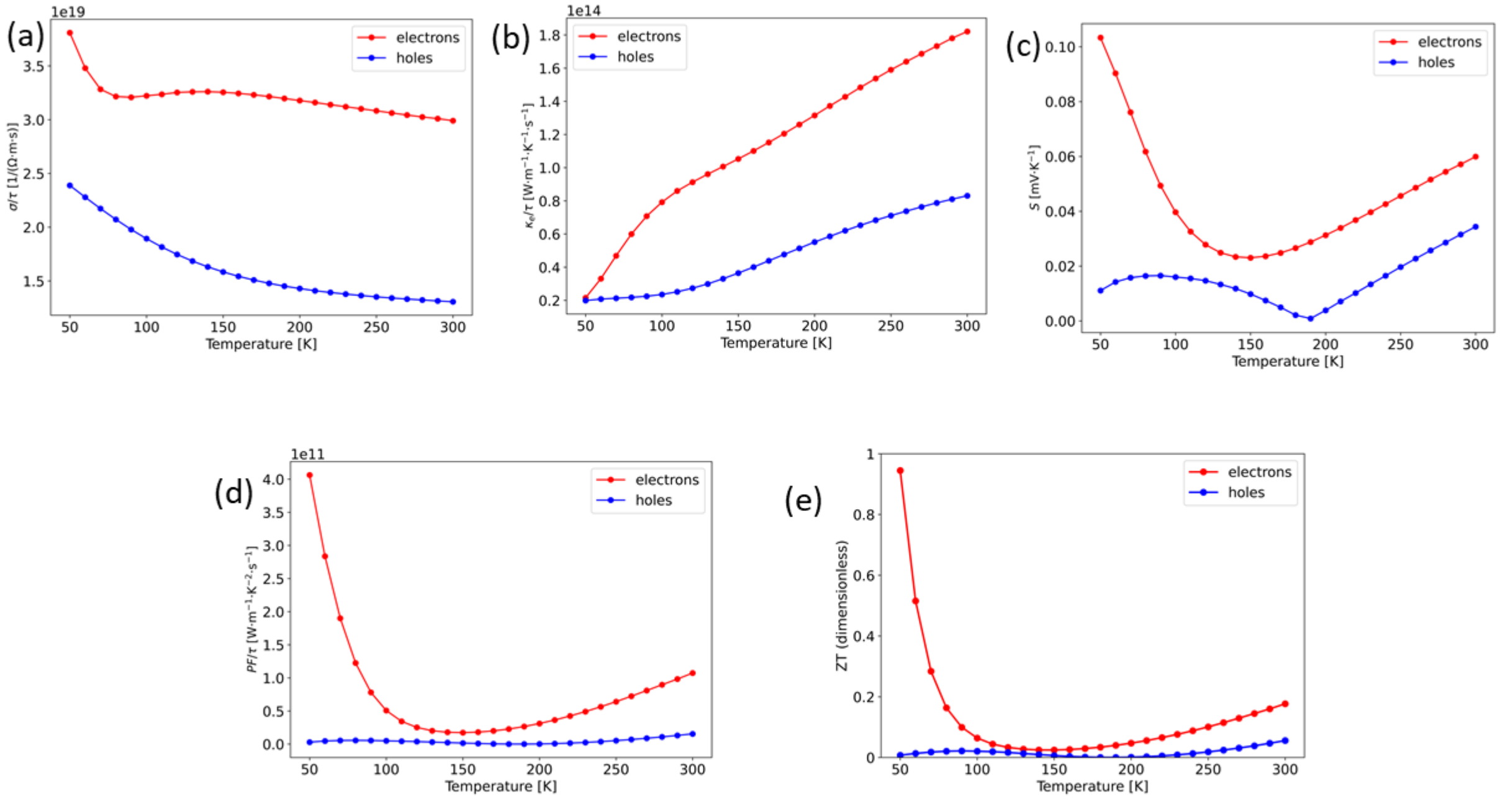

Figure 7: Temperature dependence of key thermoelectric properties in Doped condition: (a) Electrical conductivity, (b) Electronic thermal conductivity, (c) Seebeck coefficient, (d) Power factor, and (e) Figure of merit.

Fig. 7 details the transport coefficients as a function of temperature, T(K) under heavily doped conditions at a fixed chemical concentration and fixed relaxation time, constrained in Boltztrap2 code. This transition is immediately apparent in the electrical conductivity, σ [Fig. 7(a)]. that negative doping enhances electrical conductivity across all temperatures. Similar trends is observed for all of the transport coefficients [Fig. 7(a-e)] over the whole temperature window showing improved thermoelectric efficiency under negative doping. It is worth mentioning that heavy doping in the material $Mo_2SeTe_3$ demonstrates moderate lower electrical [Fig. 7(a)] and thermal conductivity [Fig. 7(b)] than the undoped condition [Fig. 6(a-b)], whereas Fig. 7(c) shows that the Seebeck coefficient is much higher for the doped condition than the undoped condition. This feature enhances the power factor [Fig. 7(d)] and the figure of merit [Fig 7(e)] alongside of the thermoelectric efficiency under doped condition. Following Fig. 7(e). The calculations reveal a striking divergence between n-type and p-type performance. For temperatures below 150 K, the n-type ZT scales aggressively, reaching an exceptional value of approximately 1 at near 50K temperature, whereas the p-type ZT remains closer to much lower value over the whole temperature domain. These data dictates a clear engineering pathway to optimize this material for real world applications, such as near-room-temperature solid-state Peltier cooling or low-grade waste heat recovery, efforts must be strictly focused on n-type doping for higher thermoelectric efficiency.

**Topological Electronic Structure in Device Performance**

The exotic electronic properties of $Mo_2SeTe_3$ motivate its use as the channel material in the proposed tunnel junction. In particular, the DFT results identify $Mo_2SeTe_3$ as a finite-gap semiconductor with SOC-induced band inversion, weak topological character, and Dirac-like surface states on symmetry-preserving surfaces. These material features are relevant to TFET design because an effective tunneling channel must satisfy two competing requirements: it must suppress OFF-state leakage through a finite band gap, while also allowing efficient source-to-channel band-to-band tunneling (BTBT) when the gate bias opens the tunneling window. The objective of this section is therefore to evaluate whether the electronic structure of $Mo_2SeTe_3$ translates into measurable device-level advantages, while clearly separating the mechanisms directly supported by TCAD simulations from those inferred from DFT-level topology.

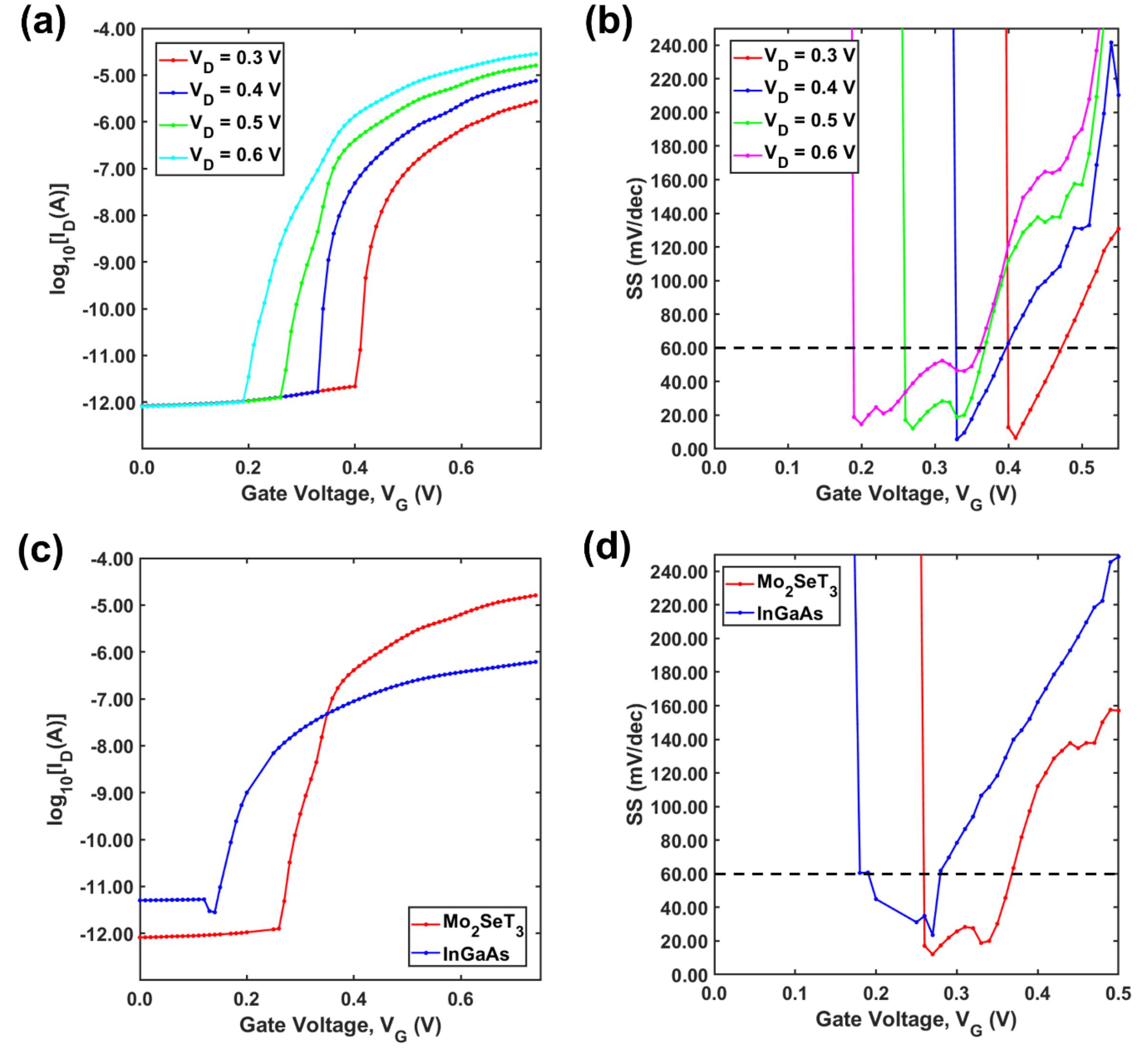

**Figure 8**. Transfer and switching characteristics of the proposed $Mo_2SeTe_3$-channel TFET**.** (a) Semilogarithmic $I_D$–$V_G$ characteristics at $V_D$= 0.3, 0.4, 0.5, and 0.6 V, showing sharp gate-controlled turn-on and increasing drain current with higher $V_D$. (b) Extracted subthreshold swing (SS), with the dashed line marking the 60 mV/dec thermionic limit. (c) Comparison of $Mo_2SeTe_3$- and InGaAs-channel TFET transfer characteristics under identical geometry, showing higher $I_{ON}$ and larger $I_{ON}/I_{OFF}$ for $Mo_2SeTe_3$. (d) SS comparison showing lower SS for the $Mo_2SeTe_3$-channel device, indicating improved gate control over the tunneling barrier.

Fig. 8(a) shows the semilogarithmic $I_DV_G$ characteristics at different drain voltages. The drain current remains strongly suppressed in the low-gate-bias region and then increases sharply as the gate voltage is raised, confirming a gate-activated switching process. It shows that $I_{ON}$ increases from 2.92 μA at $V_D = 0.3$ V to 29.8 μA at $V_D = 0.6$ V. This systematic increase with drain bias indicates that carrier injection is controlled by source-to-channel band-to-band tunneling (BTBT), rather than thermionic emission over a barrier. Fig. 8(b) shows the extracted subthreshold swing (SS) of the $Mo_2SeTe_3$-channel TFET. The average SS values are 20.9, 18.48, 21.6, and 26.2 mV $dec^{-1}$ for $V_D = 0.3$, 0.4, 0.5, and 0.6 V, respectively. All values are well below the room-temperature thermionic limit of 60 mV $dec^{-1}$, demonstrating that the device operates through tunneling-window modulation rather than thermionic carrier injection. The subthreshold swing is defined as, $SS = \frac{\partial V_G}{\partial \log_{10} I_D}$, and, for a TFET, it is governed by how efficiently the gate modulates the tunneling barrier rather than by thermal carrier emission. The low SS indicates that a small change in gate voltage produces a large change in drain current, which is a direct signature of efficient gate control over the source/channel tunneling barrier[51]. The slightly different SS values at different drain biases can be understood from the combined role of gate-induced band bending and drain-assisted tunneling-window broadening. Although a higher drain voltage improves carrier extraction and increases $I_{\mathrm{ON}}$, it may also slightly broaden the tunneling window, causing modest variation in the extracted SS.

To demonstrate the performance enhancement achieved by using $Mo_2SeTe_3$ instead of a conventional channel material, similar simulations were conducted using the same device structure and geometry, with only the channel material changed to InGaAs. Fig. 8(c) compares the transfer characteristics of the $Mo_2SeTe_3$-channel TFET with those of the InGaAs-channel TFET. Under the same device architecture, the $Mo_2SeTe_3$-channel device exhibits more than 10 times higher ON-current than the InGaAs-channel device. In addition, the $Mo_2SeTe_3$ TFET achieves a switching ratio ($I_{\mathrm{ON}}/I_{\mathrm{OFF}}$) ratio of approximately $10^7$, whereas the InGaAs-channel reference device shows a switching ratio of approximately $10^5$. This comparison provides direct device-level evidence that replacing the conventional InGaAs channel with $Mo_2SeTe_3$ improves both current drive and switching contrast. The higher $I_{\mathrm{ON}}$ indicates more efficient tunneling injection in the ON state, whereas the larger switching ratio indicates that the OFF-state leakage remains effectively suppressed before the tunneling window opens.

Fig. 8(d) compares the SS behavior of the $Mo_2SeTe_3$- and InGaAs-channel devices. The $Mo_2SeTe_3$-channel TFET exhibits a lower SS than the InGaAs-channel TFET, demonstrating more efficient gate control over the tunneling barrier. This means that, for the same change in ($V_G$), the tunneling probability increases more rapidly in the $Mo_2SeTe_3$ device. Therefore, Fig. 8(c,d) collectively establishes that $Mo_2SeTe_3$ improves the two key requirements of TFET operation: high ON-state tunneling current and steep subthreshold switching, while maintaining a low OFF-state current.

The physical origin of these trends can be understood using standard tunneling-transport theory. In a TFET, the drain current can be interpreted using a Landauer-type tunneling-current expression[53], [79]:

$$I_D = \frac{2q}{h} \int T(E) M(E) [f_S(E) - f_D(E)]\, dE$$

where T(E) is the energy-dependent tunneling transmission probability, M(E) represents the number of available tunneling modes or density-of-states contribution within the tunneling window, and $f_S(E)$ and $f_D(E)$ are the source and drain Fermi–Dirac occupation functions, respectively. This equation shows that $I_D$ increases when the gate bias enlarges the tunneling window, increases the number of available tunneling states, or enhances the transmission probability. In the present device, the sharp increase in current observed in Fig. 8(a) and the higher $I_{\mathrm{ON}}$ of $Mo_2SeTe_3$ relative to InGaAs in Fig. 8(c) are consistent with more efficient gate-induced source/channel tunneling.

The tunneling probability is exponentially sensitive to the barrier profile and can be described using the Wentzel–Kramers–Brillouin (WKB) approximation [80]:

$$T(E) \approx \exp\left[-2\int_{x_1}^{x_2} \kappa(x,E)\,,dx\right]$$

with

$$\kappa(x,E) = \frac{\sqrt{2m_r^*[U(x)-E]}}{\hbar}$$

where $m_r^*$ is the reduced tunneling effective mass, $U(x) - E$ is the local tunneling barrier energy, and $x_1$ and $x_2$ are the classical turning points. These expressions show that the drain current depends exponentially on the tunneling barrier height, tunneling width, and reduced effective mass. Therefore, even moderate improvements in source/channel band alignment, tunneling effective mass, or inter-band coupling can produce a large increase in $I_{\mathrm{ON}}$. This explains why the $Mo_2SeTe_3$-channel TFET can exhibit substantially higher ON-current than the InGaAs-channel device under identical geometry.

For a triangular tunneling barrier, the WKB expression can be simplified into a Kane-type BTBT form [80], [81]:

$$T \sim \exp\left[-\frac{4\sqrt{2m_r^*}E_g^{3/2}}{3q\hbar F}\right]$$

or, equivalently, the BTBT generation rate may be expressed as [82], [83]

$$G_{\mathrm{BTBT}} \propto F^P \exp\left(-\frac{B}{F}\right)$$

where F is the local electric field at the tunneling junction, P depends on the tunneling process, and B contains the dependence on band gap and tunneling effective mass. These relations show that stronger local electric field reduces the effective tunneling barrier and exponentially enhances BTBT generation. They also explain the lower SS observed for $Mo_2SeTe_3$ in Fig. 8(b,d), because a lower SS corresponds to a faster increase of tunneling transmission with gate voltage. The SS is defined as [51].

$$SS = \left(\frac{d\log_{10} I_D}{dV_G}\right)^{-1}$$

and, when the current is dominated by tunneling transmission,

$$SS^{-1} \propto \frac{d\log_{10} T}{dV_G}$$

Thus, the lower SS of the $Mo_2SeTe_3$-channel TFET indicates that the gate voltage modulates T(E), the source/channel electric field, and the effective tunneling length more efficiently than in the InGaAs-channel reference.

The observed leakage suppression and switching-ratio improvement can also be linked to the finite semiconducting gap of $Mo_2SeTe_3$. The DFT results show that $Mo_2SeTe_3$ possesses a finite band gap of approximately 0.7507 eV. This gap is important for TFET operation because it prevents premature source-to-channel band overlap in the OFF state, thereby suppressing leakage current. At the same time, when sufficient gate bias is applied, the band edges can be aligned to activate BTBT [84], [85]. Therefore, the finite-gap nature of $Mo_2SeTe_3$ provides a balanced condition: it suppresses OFF-state leakage while still allowing strong ON-state tunneling once the gate opens the tunneling window. This explains why the $Mo_2SeTe_3$-channel device achieves a higher ($I_{\mathrm{ON}}/I_{\mathrm{OFF}}$) ratio than the InGaAs-channel device.

Beyond the finite band gap, the SOC-modified electronic structure of $Mo_2SeTe_3$ provides a plausible material-level contribution to the enhanced BTBT current. The manuscript shows that spin–orbit coupling modifies the band-edge structure and induces orbital hybridization near the Fermi level. In interband tunneling, the transition rate depends not only on the tunneling barrier but also on the coupling between initial valence-band states and final conduction-band states. A Fermi's golden rule-type expression for inter-band generation can be written schematically as [86], [87]:

$$G_{\mathrm{BTBT}} \propto |M_{cv}|^2 D_v(E) D_c(E)$$

where $M_{cv}$ is the conduction–valence inter-band coupling matrix element, and $D_v(E)$ and $D_c(E)$ are the valence- and conduction-band densities of states. SOC-induced orbital mixing near the band edges may increase the effective overlap between valence- and conduction-band wave functions, thereby supporting stronger BTBT once the source and channel bands are aligned. This interpretation is consistent with the higher ($I_{\mathrm{ON}}$) observed in Fig. 8(c). However, because $M_{cv}$ was not explicitly extracted from the TCAD simulation, this contribution should be interpreted as a plausible material-level origin rather than a directly proven mechanism.

While Fig. 8 establishes the improved transfer characteristics and subthreshold behavior of the $Mo_2SeTe_3$-channel TFET, the internal band profile, BTBT generation, electric-field distribution, and current-density maps in Fig. 9 provide the physical evidence for the tunneling mechanism responsible for this enhancement. Fig. 9(a) shows the energy-band diagram of the proposed TFET under bias. The $Mo_2SeTe_3$ channel conduction band approaches the valence band of the $p^+$ InSb source, forming the source-to-channel tunneling window required for BTBT. In terms of the WKB expression, this band alignment lowers the effective tunneling barrier and reduces the separation between the classical turning points ($x_1$) and ($x_2$). As a result, the exponential decay factor in the tunneling probability decreases, and (T(E)) increases [80]. This provides the band-profile evidence for the sharp current increase observed in Fig. 8(a).

Fig. 9(b) shows the nonlocal BTBT generation profile. The BTBT generation is strongly localized near the source/channel junction, indicating that carrier generation occurs at the intended tunneling interface rather than throughout the channel. This localization is essential for high-performance TFET operation because it confirms that the drain current originates from gate-controlled source-to-channel tunneling[88], [89]. A localized BTBT hotspot supports high ($I_{\mathrm{ON}}$), while minimizing parasitic generation away from the tunneling junction and helping maintain low OFF-state leakage[88], [90]. Therefore, Fig. 9(b) directly supports the performance trends observed in Fig. 8(a,c).

Fig. 9(c) shows the electric-field distribution and confirms that the maximum electric field occurs at the source/channel interface. This field concentration is the electrostatic origin of enhanced tunneling. The effective tunneling length can be approximated as [91]:

$$\lambda_{\text{tun}} \approx \frac{E_g}{qF}$$

where $E_g$ is the tunneling band gap and (F) is the local electric field. This expression shows that increasing (F) reduces the tunneling length. Because the tunneling probability depends exponentially on the barrier width, the field maximum observed in Fig. 9(c) explains both the enhanced $I_{\text{ON}}$ and the lower SS of the $Mo_2SeTe_3$-channel TFET. In other words, the same electric-field localization that increases BTBT generation also enables rapid gate modulation of the tunneling barrier.

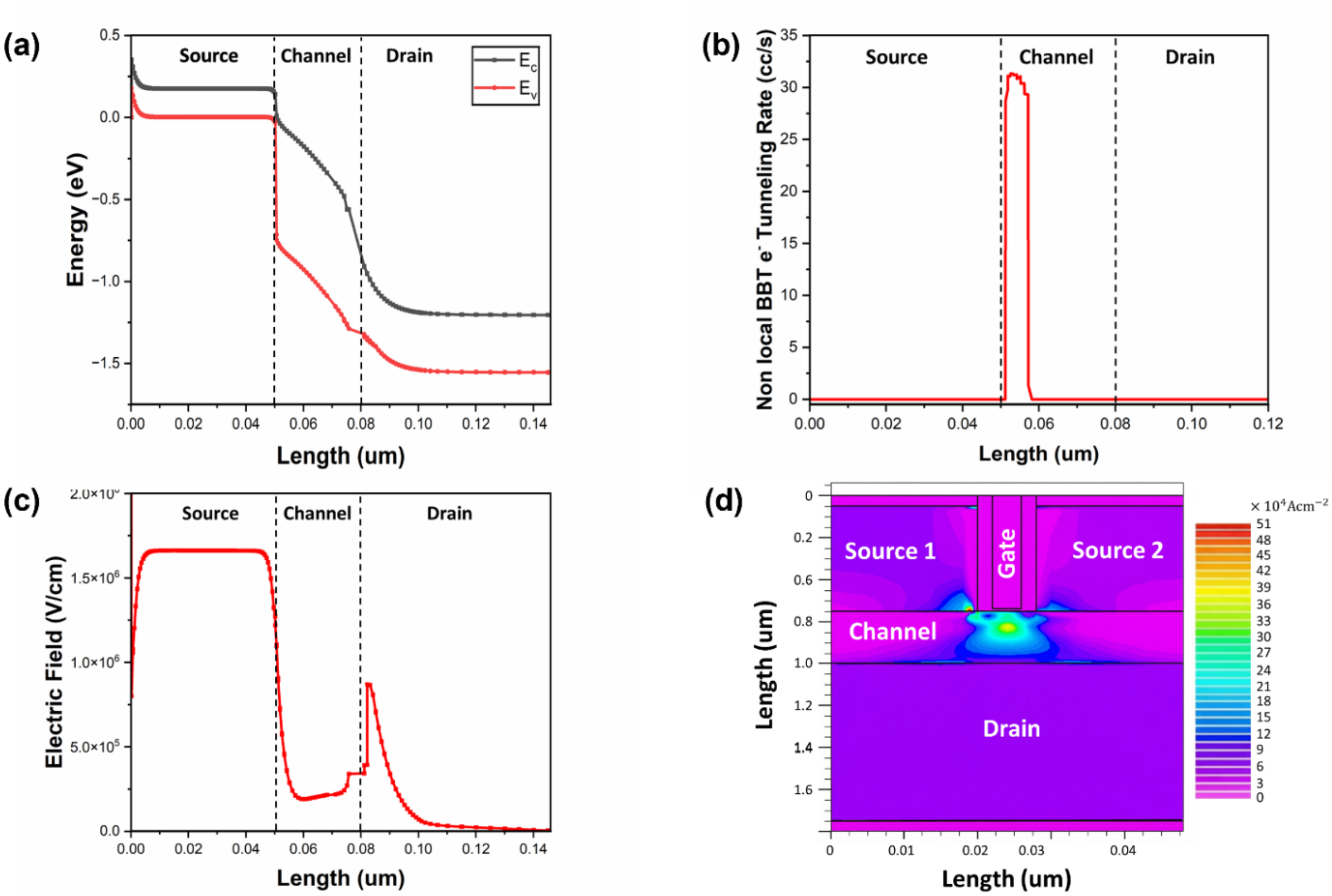

**Figure 9.** Simulated internal physical distributions of the proposed TFET at $V_D$=0.5 (a) Energy-band diagram showing the approach of the channel conduction band edge toward the source valence band edge at the source/channel interface, enabling band-to-band tunneling. (b) Nonlocal band-to-band tunneling (BTBT) rate profile, indicating strongly localized carrier generation near the tunneling junction. (c) Electric-field distribution, showing field concentration at the source/channel interface where tunneling is initiated. (d) Current-density distribution, illustrating the dominant carrier transport path through the channel region toward the drain.

Fig. 9(d) presents the current-density distribution. The current path extends from the tunneling junction through the $Mo_2SeTe_3$ channel toward the drain, confirming efficient carrier collection after BTBT generation. This result supports the interpretation that the generated carriers are transported effectively through the channel once injected. It verifies the spatial current-flow path in the simulated TFET. Hence, $Mo_2SeTe_3$ provides efficient post-tunneling carrier transport in the simulated structure, with possible additional contribution from topological boundary states only if such states are accessible under the device operating conditions.

Overall, Fig. 8 demonstrates that the $Mo_2SeTe_3$-channel TFET outperforms the InGaAs-channel reference by delivering higher, a larger ($I_{\mathrm{ON}}/I_{\mathrm{OFF}}$) ratio, and lower SS under the same device architecture. The mathematical tunneling framework shows that this improvement is consistent with stronger gate-controlled BTBT, favorable source/channel band alignment, reduced tunneling decay, and enhanced electric-field localization. The finite semiconducting gap of $Mo_2SeTe_3$ explains the suppressed OFF-state leakage, while its SOC-modified band-edge structure and orbital hybridization provide a plausible origin for stronger interband coupling during BTBT. The internal profiles in Fig. 9 confirm that the performance enhancement originates from source/channel band alignment, localized nonlocal BTBT generation, electric-field concentration, and efficient current collection. The weak topological phase and Dirac-like surface states of $Mo_2SeTe_3$ may be beneficial for post-injection transport, but their direct role in the simulated TFET performance should be treated cautiously and verified through future surface-resolved quantum-transport or experimental studies.

# Conclusion

In summary, $Mo_2SeTe_3$ is established as a weak topological semiconductor in which spin–orbit coupling drives a band inversion across a finite indirect bulk gap of ~0.75 eV, yielding topological indices (0;001) and symmetry-protected Dirac surface states confirmed by Wannier spectral calculations — providing unambiguous bulk-boundary correspondence. Phonon and elastic analyses further confirm the structural integrity necessary for scalable device integration. The material simultaneously exhibits strongly anisotropic optical response, large dielectric polarizability, and broad-spectrum plasmonic activity, alongside a competitive thermoelectric figure of merit (ZT ≈ 1) under n-type doping, demonstrating the material's strong potential for thermoelectric applications. Its mechanical resilience and favorable thermoelectric performance, particularly under carrier doping, highlight its adaptability for energy-efficient and flexible electronic architecture. Crucially, implementation of $Mo_2SeTe_3$ as the tunnel channel in a dual-source TFET architecture demonstrates that topologically enhanced interband coupling and SOC-driven orbital hybridization fundamentally amplify band-to-band tunneling efficiency, yielding subthreshold swing below the thermionic limit and a high ON/OFF current ratio — while symmetry-protected surface states inherently suppress backscattering in scaled geometries. These results collectively position $Mo_2SeTe_3$ at the intersection of topological quantum physics and energy-efficient nanoelectronics, establishing a clear and rigorous materials-to-device framework for exploiting topological semiconductors in next-generation quantum transistors, spintronic, and optoelectronic technologies.

**Acknowledgements**

The authors acknowledge Department of Physics, University of Dhaka, and Bangladesh Research and Education Network (BdREN) for computational support of this work. This manuscript was edited with the assistance of generative AI tools in accordance with the editorial policies of Springer-Nature 2025 (https://www.nature.com/nature-portfolio/editorial-policies/ai).

## Author Contributions

N.H. and A.K. conceived the project idea. Z.S.M. and S.K.A. conducted all the DFT and TCAD simulations. Data analysis, visualization, and manuscript writing were carried out by Z.S.M., and S.K.A. T.M. and A.K. provided support with computational resources. N.H. and A.K. reviewed, edited the manuscript, and supervised the project comprehensively. All authors reviewed the final version of the manuscript in its entirety.